\documentclass[11pt,twoside]{article}

\usepackage{asp2006}
\usepackage{epsf}
\usepackage{lscape}

\begin{document}
\title{Hotspots, Jets and Environments}   
\author{Martin J. Hardcastle}   
\affil{School of Physics, Astronomy and Mathematics, University of
  Hertfordshire, College Lane, Hatfield AL10 9AB, United Kingdom}    
\newcommand{\etal}{{et al.}}
\begin{abstract} 
I discuss the nature of `hotspots' and `jet knots' in the kpc-scale
structures of powerful radio galaxies and their relationship to
jet-environment interactions. I describe evidence for 
interaction between the jets of FRI sources and their local
environments, and discuss its relationship to particle acceleration, but the
main focus of the paper is the hotspots of FRIIs and on new
observational evidence on the nature of the particle acceleration
associated with them.
\end{abstract}
\section{Introduction}
In the conventional picture of powerful extragalactic radio sources,
the Fanaroff-Riley classification of an object \citep{fr74} is
determined by the physics of the interaction between the
kiloparsec-scale jets and the external medium. This is thought to take
place in two very different ways. In the low-power FRI sources there
is a good deal of evidence (see, e.g., \citealt{lb02}; Laing,
these proceedings) that the jets normally decelerate from relativistic
to sub-relativistic speeds gradually over scales of 1--10 kpc.
Deceleration requires entrainment of external material in order that
momentum should be conserved \citep[e.g.,][]{b84,k94}
and the jets on these scales are often apparently in direct contact
with the external medium, so that direct jet-medium interactions are a
possible mechanism for the deceleration. By contrast the jets in the
more powerful FRII sources remain relativistic, often out to scales of
hundreds of kpc or more, until they decelerate abruptly (i.e., on
scales that are much less than the length scale of the jet) at a shock
or shocks which involve the direct interaction of the jet not with the
external medium but with the relativistic plasma that fills the lobes
in which the jets are typically embedded, giving rise to the observed
hotspots. (For the purposes of this paper I will ignore the vast range
of intermediate cases and peculiar objects that should really be
accommodated in a scheme of this kind.)

One important {\it similarity} between the deceleration regions of the
two types of source is that both are associated with particle
acceleration. In the FRII sources the detection of optical synchrotron
emission from hotspots provided early evidence in favor of a beam
model with local particle acceleration at the hotspots (e.g.,
\citealt{mrhy89}; \citealt{myr97}; but cf.\ \citealt{gswb01}). The
current probable detection of X-ray synchrotron emission from hotspots
(e.g., \citealt{khwm05}, and see below) makes it very hard to evade the
conclusion that {\it in situ} particle acceleration at the hotspots is
necessary. In the FRIs the situation is similar: optical jet
detections meant that only models involving special geometries could
evade the necessity for {\it in situ} particle acceleration
\citep[e.g.,][]{bm93,hapr96} and these models are essentially
completely ruled out by the X-ray data \citep[e.g.,][]{hbw01}. X-ray
jet detections are common \citep{wbh01} in powerful FRI sources, so
that it seems likely that particle acceleration is always associated
with jet deceleration in FRI jets. Since we have no direct evidence
for particle acceleration anywhere else in the large-scale structure
of either class of radio source, these regions may be the place where
the electron energy spectrum observed throughout the rest of the
source is determined, and so it is important to understand the
physical processes that are going on in the jets of FRIs and the
hotspots of FRIIs, using, if possible, insights gained from each class
of source to understand the other.

In this paper my focus will be on several outstanding problems in our
understanding of the compact features of FRII sources and what we can
learn from them about jet-environment interaction, but I begin by
briefly and subjectively reviewing what we can learn about
jet-environment interaction and particle acceleration from the FRIs,
the closest of which can be studied in much more detail than is
possible for any FRII.

\section{FRI Sources}

As discussed above, the 1--10 kpc-scale jets of FRI sources have to
interact with the environment in order to decelerate, although the present
observational modelling of this process (see Laing, these proceedings)
does not allow us to extract the microphysics of the jet-environment
interaction. FRI jets should therefore be good places to look for
evidence for jet-environment interactions.

Detailed radio and optical images of FRI jets often show a `knotty'
structure with many compact features --- the best-known example is the
well-studied radio and optical jet of M87 \citep[e.g.,][]{ohc89}. The
nature of the knots is not obvious. At least some appear to affect the
whole jet (e.g., knot A in M87) which raises the possibility that they
are hydrodynamical features such as internal shocks. If this were the
case, there would be some analogy between these features and the
hotspots of FRII sources. However, bright
compact features do not seem to be present in the radio emission of
all FRI jets, and X-ray observations of jets (backed up in some cases
by optical data) often show smooth diffuse X-ray emission, which
implies distributed particle acceleration, since the X-ray loss
spatial scale (defined as the speed of light multiplied by the synchrotron loss
timescale for an electron radiating in the X-ray assuming a magnetic
field strength of the order of the equipartition value) is so short
that X-ray synchrotron emission always tells us the location of {\it
  current} particle acceleration.

The closest FRI radio galaxy, Centaurus A, provides an important
testbed for models of jet deceleration and particle acceleration in
FRI jets. Its distance (3.4 Mpc) means that the resolution of {\it
Chandra} is $\sim 10$ pc, comparable to the loss spatial scale. This
cannot be achieved in any more distant FRI source. Cen A's detailed
X-ray and radio structure (e.g., \citealt{kfjm02}; \citealt{hwkf03,hkw06}; see Hardcastle \etal , these proceedings, for images)
shows complex, knotty structure in both the radio and X-ray images,
with structure on scales that would be unresolved in any other FRI
jet. Crucially, there is a good deal of small-scale dissimilarity
between the X-ray and radio structures: there are bright radio knots
that are poor X-ray sources while many of the compact X-ray sources have only
faint or absent radio counterparts. In \citet{hwkf03} we
used the dynamical information available from multiple VLA images to
argue that the X-ray bright knots are stationary while the
radio-bright, X-ray-faint knots are moving down the jet at $\sim
0.5c$. This led us to argue that the compact X-ray bright features are
localized shocks in the fluid flow, probably as a result of
interaction with features from the external medium in the manner
proposed by \citet{bk79}. This is the first direct
association between the small-scale dynamics of an FRI jet and the
acceleration of high-energy particles.

It is important to realise, though, that the emission in compact knots
in the inner jet of Cen A represents only a fraction of the total
X-ray emission in the jet (moreover, a fraction that clearly decreases
with distance from the active nucleus). A second, diffuse component of
X-ray emission is also present, and both from the luminosity function
of knots \citep{ksat06} and from the spectral differences between
knots and diffuse emission (Hardcastle \etal , these proceedings \& in
prep.) it seems likely that this has its origin in a different
particle acceleration process. In some cases \citep[e.g.,][]{hkw06} we
may be able to associate diffuse particle acceleration with
hydrodynamical features affecting the whole jet, but in others we are
not. It seems likely that this diffuse acceleration process is related
to the bulk deceleration of the jet flow (although we do not have
evidence for bulk deceleration in Cen A) and it may well be the
dominant process in other, more distant jets. Both in Cen A and in
other jets with better-studied optical properties, the overall
radio-through-X-ray spectra are inconsistent \citep{hbw01,pw05,hkw06}
with simple continuous-injection models \citep{hm87} and the nature of
the `diffuse' particle acceleration changes as a function of distance
along the jet, with a decreasing X-ray/radio ratio and a steepening
X-ray spectral index up till the point where the X-rays disappear. We
do not as yet have any clear idea what the microphysics of this
`diffuse' acceleration process are, but it is essential to understand
it to make further progress in this area, and I will argue in the
following section that we can no longer ignore it in FRII hotspots as
well.

\section{FRII Sources}

\subsection{When Is a Knot Not a Knot?}

\begin{figure}
\epsfxsize 6.5cm
\epsfbox{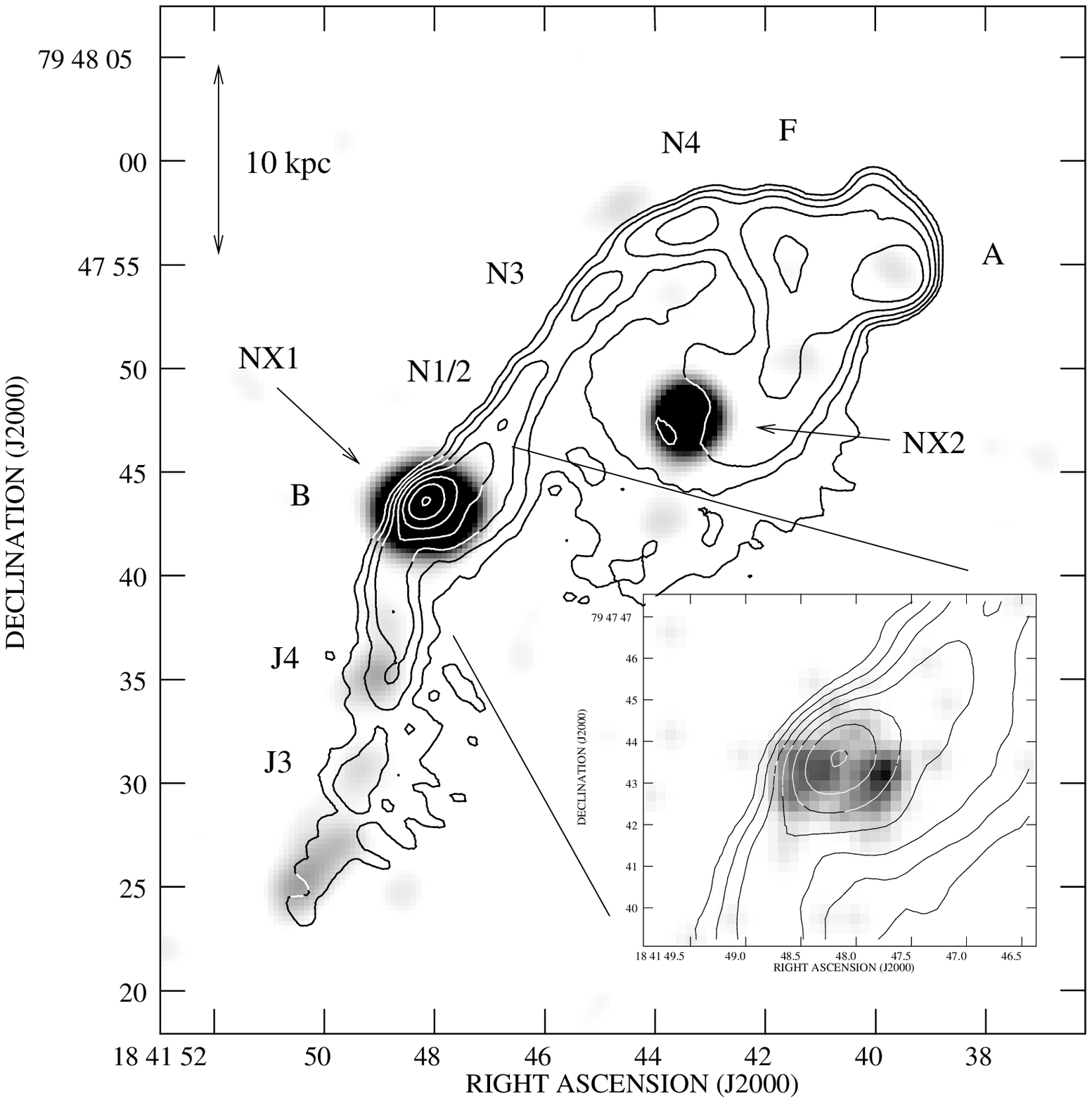}
\hskip -10pt
\epsfxsize 6.8cm
\epsfbox{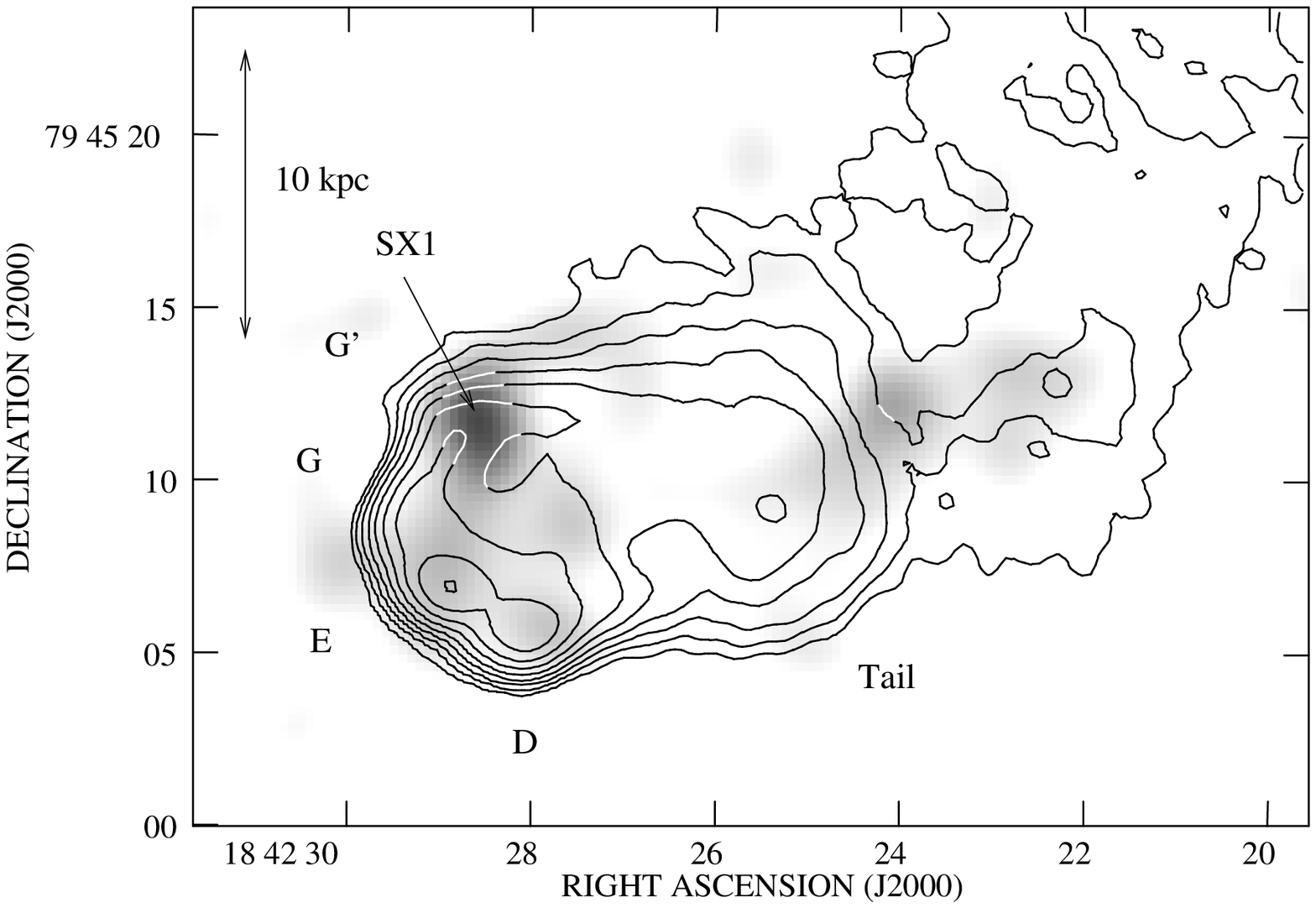}
\caption{The hotspots of 3C\,390.3 in radio and X-ray (adapted from
  \citealt{hck07b}). The greyscales show X-ray emission and the contours
  are from a 5-GHz VLA map. In the radio, note the continued
  collimated outflow after knot B (usually called the primary
  hotspot). In the X-ray, note the non-detection of the northern
  hotspot A and the diffuse emission seen from the southern hotspot.}
\label{390.3}
\end{figure}

I begin this section by making a trivial point about nomenclature that
is nevertheless important for what follows in the rest of the paper,
particularly \S\ref{knots}. This concerns when, and why, we refer to a
compact feature as a hotspot rather than a `jet knot'. It is worth
noting first of all that the definition of `hotspot' itself, as
applied to radio maps, is not widely agreed on: for example, two
widely cited papers on the detailed structure of FRII sources,
\citet{bhlb94} and \citet{lbdh97}, propose incompatible definitions.
However, all definitions generally expect the hotspots to be found at
the end of the source, since our theoretical expectation
\citep[e.g.,][]{br74} is that these will be the ones associated with
the jet termination. In particular, features that are part of a
well-collimated jet are generally classed as `jet knots'. This
division can be ambiguous --- some features near the end of jets but
with clearly continuing well-collimated outflow are nevertheless
usually referred to as hotspots (e.g., hotspot B of 3C\,390.3:
\citealt{lp95}, and Fig.\ \ref{390.3}). The existence of multiple
bright compact features associated with the jet termination (see
\S\ref{multiple}, below) is evidence that a definition of hotspots as
directly equivalent to `the location of jet termination' is
inadequate: in many places the jet does not seem to terminate at a
single location or in a single step. We know that some features that
would generally be classed as jet knots (e.g., knot F6 in 3C\,403,
\citealt{khwm05}) produce optical and X-ray emission and are thus
probably sites of particle acceleration, blurring the distinction
between jet knots and hotspots further. This ambiguity can be and has
been used in a self-serving way (the present writer is as guilty of
this as anyone else) to `define away' problems with the picture that
is being presented. A better definition would include some of the
actual physics of the jet at the location of the compact feature, but
this is hard since in many cases jets are faint or invisible --- the
`jet knot' may be the only indication that a jet is present at all. In
what follows I will use the traditional terminology, but the reader
should be aware that the term `jet knot' is thus used to encompass a
number of features with diverse characteristics and probably
significantly different physical origins.

\subsection{Jet Knots As Environmental Interactions}
\label{knots}

\begin{figure}
\epsfxsize 12cm
\epsfbox{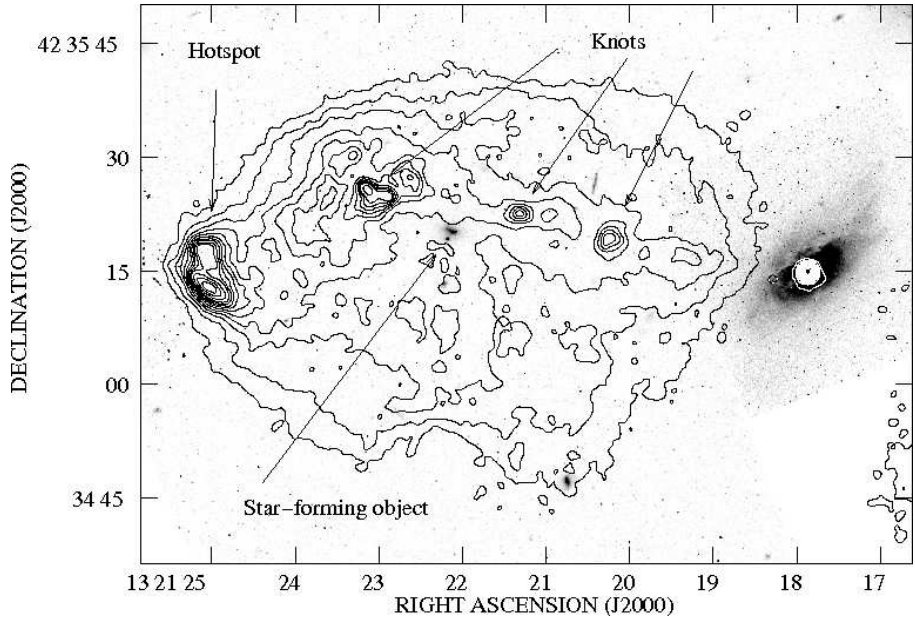}
\caption{Hotspots, jet knots and environmental interactions in the E
  lobe of 3C\,285 (from data presented by \citealt{hkwc07a}). Greyscale
  shows {\it HST} image, contours are from an L-band VLA image.}
\label{285}
\end{figure}

Is there a class of jet knots in FRIIs that are related to direct
environmental interactions with the jet, as we have argued in the case
of Centaurus A? One place to look for such jet knots would be in
sources where there is an independent suggestion of a jet/external
medium interaction in the shape of `jet-induced star formation'.
Well-studied candidates for this process include 3C\,285 \citep{vd93}
and 3C\,34 \citep{blr97}. However, there is little evidence in any of
these sources that the interaction, if present, affects the jet
itself. In 3C\,285, for example, the images of \citeauthor{vd93} show
the jet to contain several knots, but none is spatially coincident
with the position of the star-forming object that is believed to be
interacting with the jet (Fig.\ \ref{285}). Thus not only are the
knots in the jet not apparently related to environmental interactions,
but the proposed environmental interactions do not apparently produce
knots.

One FRII source where a jet-environment interaction apparently does
affect the jet is 3C\,321 (Evans \etal , ApJ submitted). Here an
otherwise normal FRII source has a jet that is disrupted downstream of
a close passage between the jet and a companion galaxy, forming an
FRI-like jet structure on one side. The clear compact hotspots on the
same side suggest that this is a temporary interaction that has been
going on for only $\sim 10^6$ years. Immediately adjacent to the
companion galaxy is a compact jet knot. However, this knot is
relatively faint and is not a source of any detectable optical or
X-ray emission, unlike some other jet knots in FRIIs: nor is the
downstream `FRI-like' jet an X-ray or optical source. The analogy
between this source and the jet-medium interactions seen in FRIs is
thus not particularly strong.

The lack of any strong relationship between jet knots and
environmental interactions lends support to a picture in which jet
knots in FRIIs have multiple possible origins. The regularly or
quasi-regularly spaced knots seen in many jets \citep[see,
e.g.,][]{bhlb94} seem most likely to be related to internal shocks in
the jet or to fluid dynamical instabilities.

\subsection{Hotspots and the Jet-External Medium Interaction}

Can the features conventionally classed as hotspots give us
information about jet-environment interactions? Since, as discussed
above, the `environment' of the jet at the jet termination point is
the lobe or cocoon, it is clear that hotspots can in principle tell us
about interactions with the lobe plasma: I discuss this in
\S\ref{multiple}. Here, however, we can ask whether hotspots tell us
anything about the {\it external} environment.

We expect that the jet should terminate close to a boundary of the
lobe or plume that the jet inhabits: the boundary is, of course, the
location of the contact discontinuity between the external medium and
the relativistic plasma that has previously passed through the jet. A
consequence of this is that we do not expect to see hotspots that are
physically in the centre of the lobe. A true hotspot (as opposed to a
jet knot) seen at a distance from any lobe boundary is assumed to be
there as a result of projection of an interaction between the jet and
the front or back boundary of the lobe. A nice example of the jet/lobe
boundary interaction is given by the jet termination in wide-angle
tail radio galaxies \citep{h99,hs04}:
in these sources, which we argue are similar to FRII
radio galaxies with peculiarly-shaped lobes, compact hotspots are
sometimes present and sometimes absent, and it is plausible that they
are present only when the geometry of the plume and the current
direction of the jet are such that the jet intersects the edge of the
plume. Here the interaction between the external environment and the
`cocoon' plasma determines where the contact discontinuity is, and the
intersection of the jet and the contact discontinuity, if present,
determines where the hotspot is --- giving an indirect but important
relationship between the hotspot and the external environment. Other
examples are provided by the `bottle-neck' hotspot structures seen in
some FRII radio galaxies, particularly at low radio luminosities
\citep[see, e.g.,][]{lbdh97},
which suggest an interaction between the lobe and a low-density region
of the external medium. In general, though, while the {\it location}
of hotspots can tell us something about the location and nature of the
external medium, the detailed properties of hotspots do not.

\subsection{Hotspots and the Jet-Environment Interaction in the Lobes}

\label{multiple}
What then do the detailed properties of hotspots tell us? I shall
argue in this section that their complex structure and particle
acceleration behavior are telling us about the nature of the interaction
between the jet and its environment, the cocoon.

The two traditional tools for study of hotspots have been
monochromatic high-resolution (polarimetric) imaging, usually in the
GHz-frequency radio, and broad-band spectral energy distributions
usually based on data of comparatively low resolution.

The first of these shows us that hotspots are not structurally simple:
the terminations of jets rarely resemble the idealized axisymmetric
structures seen in early (and, regrettably, some current) analytical
and numerical modelling. This first became clear in the early days of
high-resolution radio imaging when multiple hotspots started to be
routinely detected in a given lobe \citep[e.g.,][]{l82}. Observationally
we try to distinguish between the most compact, `primary' hotspot in a
lobe and the more diffuse `secondary' hotspot or hotspots, though
sometimes this distinction is ambiguous. \citet{lbdh97}
introduced the useful concept of a `hotspot complex' which covers a
group of hotspots and the high-brightness emission that surrounds
them. Models for this complexity fall into two basic classes: those
that seek to preserve a single true termination for the jet, such as the still
popular dentist's drill model of \citet{s82} or the
disconnected-jet variant of \citet{cgs91}, or those that propose
that both the primary and secondary hotspot have a connection to the
energy supply, so that the jet termination takes place in stages at
the various hotspots, such as the splatter-spot model of \citet{wg85} or the jet-deflection model of \citet{lb86}.

On the other hand, the radio through optical spectral energy
distributions of hotspots (primary and secondary) have often been
found to be in agreement with relatively simple models of first-order
Fermi particle acceleration at strong shocks (e.g., Meisenheimer
\etal\ 1989), although even just based on SEDs some discrepancies were
apparent (see \citealt{myr97}, in particular the prescient closing
paragraph) while the optically resolved synchrotron emission seen in
some nearby hotspots also points to a distributed acceleration
mechanism \citep[e.g.,][]{pbm02}. Inclusion of X-ray information
in principle would give important information about the location and
nature of particle acceleration to the highest observable energies, as
in the FRI jets. However, the use of hotspot X-rays has been limited
until recently because of uncertainty about the extent to which
inverse-Compton emission is responsible for the observed X-rays. In
\citet{hhwb04} we argued that inverse-Compton emission
dominates only in the highest-luminosity hotspots, and that
synchrotron emission is the only emission process that can produce
detectable X-rays in the hotspots with the lowest radio luminosities
(i.e., in the least powerful sources). Low-power FRII sources are
therefore the best places to search for synchrotron X-rays that trace
the locations of high-energy particle acceleration.
\begin{figure}
\epsfxsize 13cm
\epsfbox{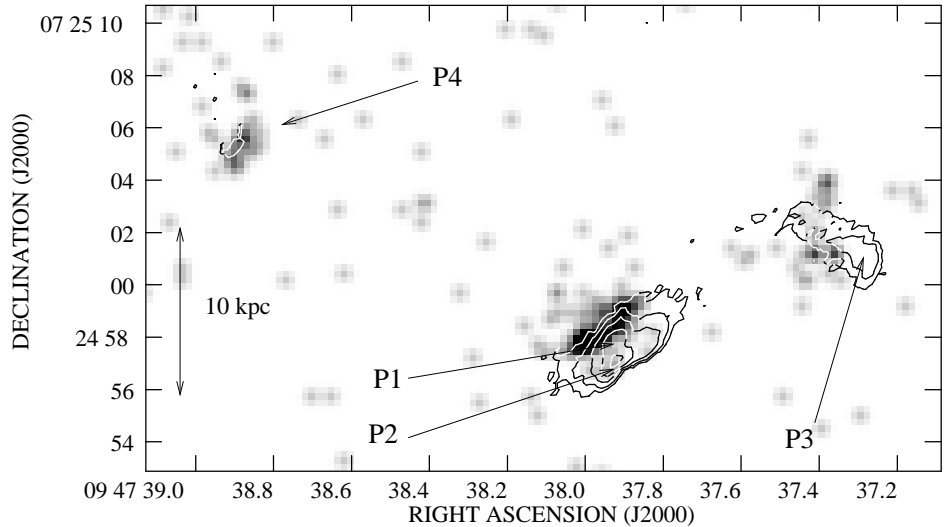}
\caption{Radio and X-ray emission from the double W hotspot of 3C\,227.
  Contours are from an 8.3-GHz radio map with $0.37 \times 0.22$
  arcsec resolution. Grayscale shows the {\it Chandra} X-ray data.
  Note the detection of X-rays associated with the `jet knot' P4, the
  primary hotspot P1/2 and the secondary P3, as well as the clear
  kpc-scale offsets between the radio and X-ray peaks in the primary
  and secondary. Image from \citet{hck07b}.}
\label{227}
\end{figure}
 
Our work in this area has been presented by \citet{khwm05}, \citet{hc05},
\citet{kbhe07} and most recently \citet{hck07b}. The key results can be
summarized as follows:
\begin{itemize}
\item X-ray emission associated with compact hotspots is common: the
  SEDs are usually consistent with a synchrotron origin for the X-rays
  and inconsistent with inverse-Compton models. There are often,
  though not always, significant offsets on $\sim 1$ kpc scales
  between the peak of the emission in the X-ray and the peak in the
  X-ray, which is not expected in a simple synchrotron/shock
  acceleration model (e.g., Figs \ref{390.3} and \ref{227}: see also
  Erlund \etal , these proceedings).
\item X-ray emission is also often seen from diffuse regions around or
  behind the hotspots, with size scales $>10$ kpc (e.g., Fig.\
  \ref{390.3}). If this X-ray emission is also synchrotron --- and no
  other model is at all plausible --- then distributed particle
  acceleration must be taking place on these scales. This is very
  plausibly related to the distributed acceleration mechanism required
  in the FRI jets.
\item Some secondary hotspots are X-ray sources and so inferred to be
  sites of high-energy particle acceleration (e.g., Fig.\ \ref{227}).
  This implies that at least some secondary hotspots have
  ongoing access to a supply of energy, and are thus {\it not}
  produced by the simple dentist's drill mechanism. Other hotspots are
  not detected in the X-ray and are therefore significantly less
  efficient at accelerating particles to the energies required for
  X-ray emission. We find that all X-ray-emitting
  hotspots are compact, but not all compact hotspots are X-ray
  sources.
\end{itemize}

We interpret these complex hotspot behaviors in the context of models
in which the hotspots are transient features created by the
interaction of the lobe fluid with the jet. This process has been seen
in many detailed simulations of light jets in which the assumption of
axisymmetry has been removed \citep[e.g.,][]{n96}. The jet/lobe
interaction causes the jet to move about in the head of the source on
scales much less than the source lifetime. In this picture individual
hotspots are telling us more about `weather' in the lobe than about
source physics: in particular, we cannot hope to understand the
mechanisms of multiple hotspot formation by looking at any individual
source, since the simulation shows that all the traditional processes
may be capable of operating at different times. Simulations that trace
shocks and particle acceleration also give important clues to the
interpretation of the X-ray emission. \citet{tjr01} carried out
three-dimensional MHD simulations that modeled the transport of
relativistic electrons and of particle acceleration at shocks. They
found that the interaction of the jet and the backflowing plasma at
the head of the jet produced what they called a `shock-web complex',
``a region of shocks of varying strengths and sizes spread throughout
the source''. Even when there was a simple terminal shock, not all the
jet material necessarily passed through it, and the terminal shock was
not always the strongest shock in the system. While it is not clear
that their simulations are perfectly matched to real radio sources,
they are capable of producing simulated synchrotron images that show
apparent clear discrete multiple hotspots \citep{tjrp02} and in these
cases the particle acceleration is not necessarily well matched to the
locations of the hotspots: hotspot locations in their model can have
more to do with magnetic field amplification than with particle
acceleration. The notion of a `shock-web complex' at the head of the
jet could help to explain the diffuse X-ray emission now seen in the
radio-bright but non-compact source head regions of a number of
objects, as discussed above, while the idea that the particle
acceleration region may not always be co-spatial with the observed
radio hotspot might help to explain observed offsets.

It seems very likely therefore that hotspots are indeed telling us
about jet-environment interactions, but that the environment with
which they are interacting is the cocoon, whose properties we cannot
probe directly. Progress in this area may come from further numerical
modelling or possibly from a better understanding of the nature of the
cocoon plasma and the distribution of particles and magnetic field
within it.

\section{Summary}

The key results from the work described in this paper can be
summarized as follows:

\begin{itemize}
\item In FRI jets, there is some direct evidence that some jet
knots are due to environmental interaction, and we
have an idea of the mechanism by which the interaction produces the
knots. However, localized interactions are most likely not responsible
for most of the high-energy particle acceleration in these jets.
\item In most FRII jets there is little evidence that knots
know about the environments, but there are
counterexamples. `Jet knots' in FRII jets are a heterogeneous class
and the terminology should be used cautiously.
\item The detailed properties of hotspots in FRIIs are
predominantly a result of interaction between the jet
and the cocoon plasma, and there is much still to be
understood about how this works. However, simple models of the locations and
mechanisms of particle acceleration in these objects are now ruled out
by a variety of observations.
\end{itemize}

\acknowledgements

I thank my collaborators past and present who have worked
with me on hotspot and jet radio X-ray observations, particularly
Ralph Kraft, Dan Harris, Diana Worrall, Mark Birkinshaw, Judith
Croston and Dan Evans. I also acknowledge generous financial support
from the Royal Society.


\end{document}